# Antiferromagnetic Order Induced by an Applied Magnetic Field in a High-Temperature Superconductor

B. Lake[1], H.M. Rønnow[2], N.B. Christensen[3], G. Aeppli[4,3], K. Lefmann[3], D.F. McMorrow[3], P. Vorderwisch[5], P. Smeibidl[5], N. Mangkorntong[6], T. Sasagawa[6], M. Nohara[6], H. Takagi[6], T.E. Mason[7].

[1]Oak Ridge National Laboratory, P.O. Box 2008 MS 6430, Oak Ridge, TN 37831-6430, U.S.A.
[2]CEA (MDN/SPSMS/DRFMC), 17 Ave. des Martyrs, 38054 Grenoble cedex 9, France.
[3]Materials Research Department, Risø National Laboratory, 4000 Roskilde, Denmark.
[4]NEC Research Institute, 4 Independence Way, Princeton, New Jersey 08540-6634, USA.
[5]BENSC, Hahn-Meitner Institut, Glienicker Strasse 100, 14109 Berlin, Germany.
[6]Department of Advanced Materials Science, Graduate School of Frontier Sciences, University of Tokyo, Hongo 7-3-1, Bunkyo-ku, Tokyo 113-8656, Japan.
[7]Experimental Facilities Division, Spallation Neutron Source, 701 Scarboro Road, Oak Ridge, TN 37830, U.S.A.

{PACS numbers: 74.72.Dn, 61.12.Ex Hk, 74.25.Ha}

One view of the cuprate high-transition temperature (high-$T_c$) superconductors is that they are conventional superconductors where the pairing occurs between weakly interacting quasiparticles, which stand in one-to-one correspondence with the electrons in ordinary metals – although the theory has to be pushed to its limit [1]. An alternative view is that the electrons organize into collective textures (e.g. charge and spin stripes) which cannot be mapped onto the electrons in ordinary metals. The phase diagram, a complex function of various parameters (temperature, doping and magnetic field), should then be approached using quantum field theories of objects such as textures and strings, rather than point-like electrons [2,3,4,5,6]. In an external magnetic field, magnetic flux penetrates type-II superconductors via vortices, each carrying one flux quantum [7]. The vortices form lattices of resistive material embedded in the non-resistive superconductor and can reveal the nature of the ground state – e.g. a conventional metal or an ordered, striped phase - which would have appeared had superconductivity not intervened. Knowledge of this ground state clearly provides the most appropriate starting point for a pairing theory. Here we report that for one high-$T_c$ superconductor, the applied field which imposes the vortex lattice, also induces antiferromagnetic order. Ordinary quasiparticle pictures cannot account for the nearly field-independent antiferromagnetic transition temperature revealed by our measurements.

La$_{2-x}$Sr$_x$CuO$_4$, is the simplest high-$T_c$ superconductor. The undoped compound is an insulating antiferromagnet, where the spin moments on adjacent Cu$^{2+}$ ions are antiparallel [8]. Introduction of charge carriers via Sr doping reduces the ordered moment until it vanishes at $x<0.13$. In addition, for $x>0.05$ the commensurate antiferromagnetism is replaced by incommensurate order [2,3,9,10], where the repeat distance for the pattern of ordered moments is substantially larger than the spacing between neighbouring copper ions. La$_{2-x}$Sr$_x$CuO$_4$ becomes a superconductor for Sr dopings of $0.06<x<0.25$; and for underdoped compounds ($0.06<x<0.13$), both superconductivity and magnetic order are present. Figure 1a shows the electrical resistance as a function of temperature $T$ and magnetic field $H$, applied perpendicular to the superconducting $CuO_2$ planes, for a sample with $x=0.10$. The resistance decreases rapidly below the zero-field transition temperature $T_c(H=0T)=29K$, and becomes zero below the irreversibility temperature $T_{irr}(H)$. The white circles locate $T_{irr}$ and show that it is a rapid function of applied field, so that even for fields much smaller than the upper critical field ($H_{c2}(x=0.10)\sim45T$ [11]), perfect conductivity does not occur until the temperature is well below $T_c(H=0T)$.

In our experiments we used magnetic neutron diffraction to measure the spin ordering in single crystals of underdoped $La_{2-x}Sr_xCuO_4$ ($x=0.10$). The technique is analogous to mapping the positions of atoms in crystals using X-ray or neutron diffraction. The superconducting $CuO_2$ planes were aligned in the scattering plane and the magnetic field was applied perpendicular to these planes. The inset in Fig. 2a shows the reciprocal space region over which data were collected. In zero field (Fig. 2a), incommensurate elastic peaks occur in the superconducting phase at low temperatures. The peaks are resolution limited, implying an in-plane correlation length of $\zeta>400Å$, similar to that observed by Kimura *et al*. [10] for $x=0.12$. The peak amplitude decreases as $T$ is increased and is entirely absent at $T=30K$. Our new finding is that external magnetic fields dramatically increase the low temperature signal. Figure 2b shows that for $H=14.5T$ and $T=2K$ the signal is three times larger than in zero field, while at 30K, just above $T_c(H=0T)$, there is very little field-induced signal. The field-induced signal is resolution limited and the magnetic in-plane correlation length ($\zeta>400Å$), is much greater than the superconducting coherence length ($\xi\sim20Å$). Since the superconducting coherence length gives the size of the vortices the magnetism cannot reside in the vortex cores alone, but must extend beyond into the superconducting regions of the material This result is important because, taken together with the large zero-frequency antiferromagnetic susceptibility of superconducting $La_{2-x}Sr_xCuO_4$ [12], it implies that as long as the vortex array is not radically different from a conventional Abrikosov lattice, superconductivity and antiferromagnetism co-exist throughout the bulk of the material.

We collected scans similar to those shown in Fig. 2 for a variety of fields and temperatures. The neutron signal is normalized using a standard phonon-based calibration to yield the ordered spin moment squared as displayed in Fig. 1b. Ordering develops just below the zero-field transition temperature $T_c(H=0T)$ and increases with decreasing temperature and increasing field. Figure 3a shows the temperature dependence; the zero field ordering (blue circles) increases gradually below $T_c(H=0T)$, reaching a maximum of $0.15\pm0.01\mu_B/Cu^{2+}$. This value was somewhat larger than the $0.10\pm0.04\mu_B/Cu^{2+}$ deduced from neutron diffraction measurements for a sample with $x=0.12$ [13]. The field-induced order for $H=5T$, 10T and 14.5T is also plotted; it was obtained by subtracting the zero field signal from the signal measured in field. The temperature dependence of the field-induced signal is clearly different from that for $H=0T$. It increases more rapidly just below $T_c(H=0T)$ and saturates for $T<10K$, where in zero field the moment is still evolving. Figure 3b shows the $H$-dependence of the field-induced order at base temperature ($T=1.9K$). The ordering increases rapidly with small fields probably reflecting the linear dependence of vortex density with field, the rate of increase slows for higher fields as the antiferromagnetic regions start to merge.

Our work represents a major departure from two previous experiments on $La_{2-x}Sr_xCuO_4$ in magnetic fields. In the first experiment, the static magnetism in a sample with $x=0.12$ was found to undergo an enhancement in an applied magnetic field consistent with our findings [14]. However this measurement was confined to a single field (10T) and temperature (4.2K) and the sample was atypical with an anomalously low value of $T_c(H=0T)=12K$. The zero-field Néel temperature, $T_N$=25K, was well above $T_c(H$=0T), and the onset temperature for the field-induced enhancement was unknown. The second experiment focused on magnetic dynamics in optimally doped La$_{2-x}$Sr$_x$CuO$_4$ ($x$=0.163) [15]. This material shows no static magnetism in either zero or non-zero fields up to 14.5T. However an applied field has the effect of recovering some of the magnetic fluctuations below the 'spin gap' which are suppressed by superconductivity in zero field. An important feature of the experiment is that the antiferromagnetic correlations develop at a temperature $T_{irr}$, i.e. they only occur once phase coherent superconductivity has been attained. Thus this result is in many ways the dual of that for our present 10% sample where the static antiferromagnetism arises first and phase coherent superconductivity is established at a lower temperature, when the ordered moments have saturated near their base temperature values.

We believe that the field-induced signal is intrinsic and arises from fundamental physical processes, even though the zero field signal may well be stabilized by the disorder inherent in a random alloy such as La$_{2-x}$Sr$_x$CuO$_4$. Thus, the magnetic field does not simply enhance the existing antiferromagnetism but induces magnetic order in its own right. There are several compelling reasons to support our belief. First, the sample is of very high quality as shown by specific heat measurements [16] and our transport data (Fig. 1a). Second, a different and poorer quality sample, also with $x$=0.10, which displays static antiferromagnetism in zero field above as well as below $T_c(H$=0T), showed field-induced ordering with exactly the same temperature dependence as reported here [17]. Third, the degree of ordering is substantial, corresponding to a volume fraction of ~50% for the incommensurate magnetic state, assuming that the ordered moment is 0.4$\mu_B$/Cu$^{2+}$, as suggested by muon spin relaxation data [18]. Finally while the zero-field signal indeed displays the gradual rise typically associated with defect-induced magnetism in correlated Fermi systems [19], the field-induced signal increases according to classical mean-field theory, suggesting a different and intrinsic mechanism (Fig. 3a).

Our data bear on two popular descriptions of the cuprates. The first states that the cuprates are describable in the same terms as ordinary solids, where small variations of key parameters lead to Fermi surface instabilities corresponding to superconductivity and insulating 'striped' phases. Within such an interpretation, stripe order is simply tuned by doping to appear because of Fermi surface nesting and lattice anomalies near $x$=1/8. However, what we have found is that below $x$=1/8, very modest (on the scale of both electron and phonon energies) fields induce antiferromagnetic order with a field-independent onset temperature and a strongly field-dependent amplitude. These results cannot follow from a simple Fermi 'quasiparticle' description, which anticipates that the Néel temperature will rise significantly as the base temperature ordered moment increases.

A second description of high-$T_c$ superconductivity does not focus on microscopic detail unlike the quasiparticle-based theories, but is instead based on the hypothesis that the cuprates possess hidden 'SO(5)' symmetry where magnetic and superconducting order parameters can be 'rotated' into each other. Calculations based on this hypothesis predict antiferromagnetic vortices in a magnetic field and have had some success in modeling the data [20-23]. The original work was performed assuming that the symmetry is exact [20]. In this scenario the antiferromagnetism is confined to the vortex cores which have size ~$\xi$=20Å, and there are no regions of the compound where superconductivity and antiferromagnetism exist simultaneously. Our data suggest that the antiferromagnetic regions are much greater than this ($\zeta$>400Å) and extend well beyond the vortex cores into the superconducting areas. Alternatively there could be interactions between the vortices that lead to coherent long range antiferromagnetic order without the magnetism spreading beyond the cores; but then the corresponding Néel temperature would be a strong function of the intervortex spacing, controlled by *H*, whereas our onset temperature is approximately field-independent.

More recent computations motivated by our experiments have relaxed SO(5) symmetry to allow co-existence of superconductivity and magnetism with the antiferromagnetic correlations extending beyond the vortex core into the bulk of the crystal [22,23]. While the underlying continuum approximations may be called into question at high fields, such a model [23] gives quantitative results in good agreement with our $T$=0 data, as shown in fig. 3b. The physical insight here is that the magnetism of the vortex state manifests a magnetic quantum critical point very close to the superconductor in the phase diagram.

Where do our discoveries leave our wider understanding of high-$T_c$ superconductivity? First, they provide the first clear evidence for intrinsic antiferromagnetism coexisting with superconductivity in the same sample. Because the relative amplitudes of the two types of order can be continuously tuned via magnetic field for fixed sample rather than via doping, experiments to examine the tradeoffs between magnetism and superconductivity will become much more detailed and reliable. Second, our data taken together with resistivity measurements [24] strongly suggest that in a magnetic field large enough to destroy superconductivity ($H>H_{c2}$), an underdoped cuprate would become an incommensurate, antiferromagnetic insulator. This means that if we want to construct a $T$=0 theory of superconductivity in underdoped cuprates, the most obvious starting point is this antiferromagnetic insulator, and not - as in conventional Bardeen-Cooper-Schrieffer-inspired theories - a metallic Fermi liquid of weakly interacting quasiparticles.

**Acknowledgments**

We thank P. Dai, P. Hedegård, S. Kivelson, H. Mook, J. Zaanen, S. Sachdev and S-C. Zhang for useful discussions. Oak Ridge National Laboratory is managed by UT-Battelle, LLC, for the US Department of Energy. H.M. Rønnow holds a Marie Curie Fellowship funded by the European Community.

Correspondence and requests for material should be addressed to B. Lake (e-mail: acilake@yahoo.com)

**Caption 1**
Magneto-transport and neutron diffraction data for $La_{2-x}Sr_xCuO_4$ as a function of temperature and magnetic field. (a) shows magneto-transport measurements parallel to the $CuO_2$ planes, obtained via a standard four-probe method; the colours indicate the electrical resistivity. In a magnetic field, the sharp transition from normal to superconducting states is broadened into a crossover region and vortices are thought to form at temperatures where the resistivity falls below its value at $T_c(H$=0T$)$. Phase coherent superconductivity, characterized by zero resistance, sets in at the much lower `irreversibility' temperature ($T_{irr}(H)$), marked by the white circles. (b) shows the square of the ordered spin moment per $Cu^{2+}$ ion as a function of temperature and applied magnetic field. The ordered moment squared is proportional to the observed neutron scattering signal and was deduced from scans similar to those shown in Fig. 2. It first becomes significant below the zero-field superconducting transition temperature ($T_c(H=0T)$), and increases with decreasing temperature and increasing field. The crystals used for these measurements were grown in an optical image furnace.

**Caption 2**
Magnetic neutron diffraction data for $La_{2-x}Sr_xCuO_4$ with $x=0.10$. The inset shows the relevant reciprocal space, labeled using the two-dimensional notation appropriate for the superconducting $CuO_2$ planes. The black dot at (0.5,0.5) represents the Bragg point associated with the commensurate antiferromagnetism of the insulating $x=0$ parent compound. The incommensurate antiferromagnetic order in our metallic $x=0.10$ material, gives rise to diffraction at the quartet of blue dots [10]. The pink line shows the trajectory of a typical scan which passes through the incommensurate peak at (0.5,0.6175) and the direction of the applied field is shown to be perpendicular to the $CuO_2$ planes. (a) gives the data collected in zero field. The red circles show that signal is present in the superconducting phase at $T=1.9K$. However, just above the superconducting transition temperature at $T=30K$, the signal has completely disappeared (blue circles). (b) shows the same scan as in (a), measured at the same temperatures but in a field of $H=14.5T$; the low temperature signal is a factor three larger than the zero field signal. The data were collected using the V2/FLEX neutron scattering spectrometer at the BER II reactor, Hahn-Meitner Institute, Berlin. The instrument was used with a collimation of guide-60'-60'-open and a PG monochromator and analyser. The energy was fixed at 7.5meV and a tuneable PG filter was used to eliminate second order scattering.

**Caption 3**
The temperature and field-dependence of the ordered spin moment squared. The data have been calibrated using a transverse acoustic phonon ($q=(1+\varepsilon,1-\varepsilon)$, $E=2meV$, sound velocity=26.9meV-Å) and are presented in units of $\mu_B^2/Cu^{2+}$. (a) shows the temperature-dependence. The zero-field signal (blue circles) increases gradually below $T_c(H=0T)$ and the dashed line is a guide to the eye. Also plotted is the field-induced signal (equal to the signal measured in field minus the zero-field signal) for fields of $H=5T$ (yellow circles), 10T (green circles) and 14.5T (red circles). The solid lines through the data are fits to mean field theory [25]. (b) shows the field-induced signal as a function of field at $T=2K$ (magenta circles). The solid line is a fit to the data of the expression, $M^2(H/H_{c2})\ln(H_{c2}/H)$ deduced by Demler *et al*. [23], with fitted parameter $M^2=0.12\mu_B^2/Cu^{2+}$.

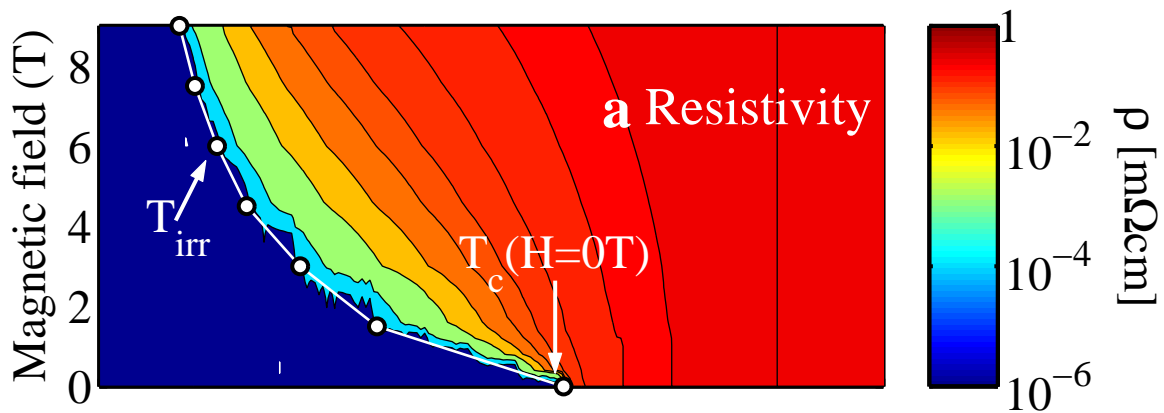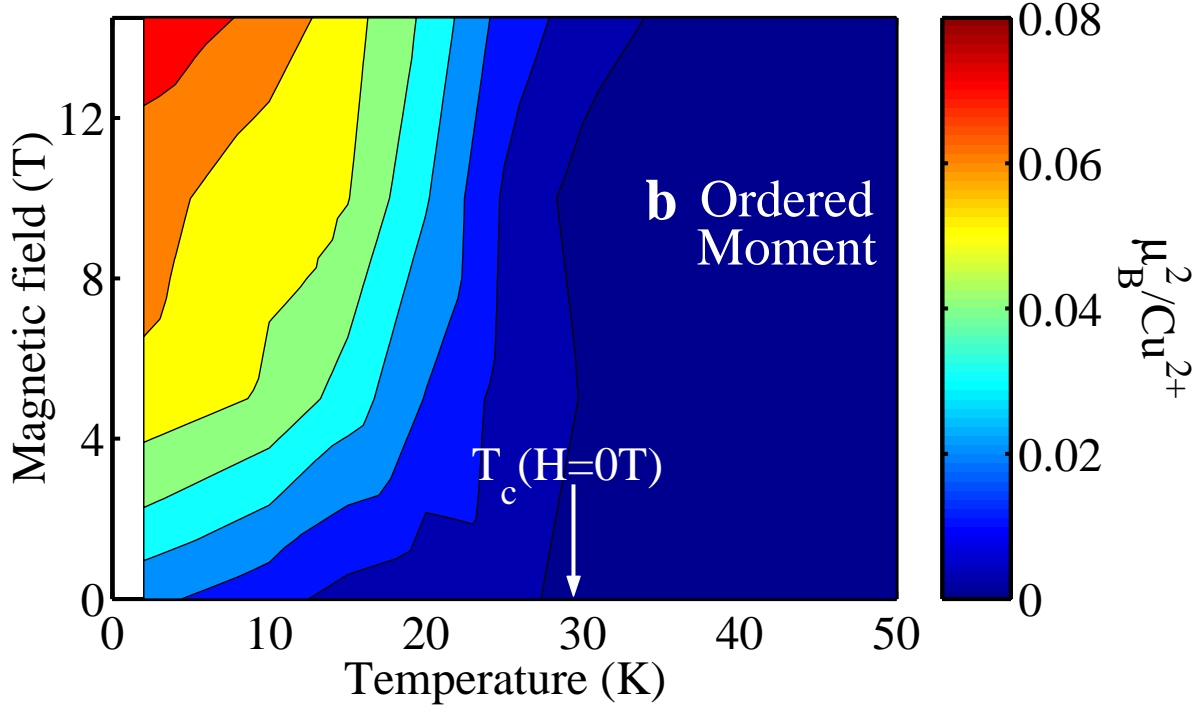

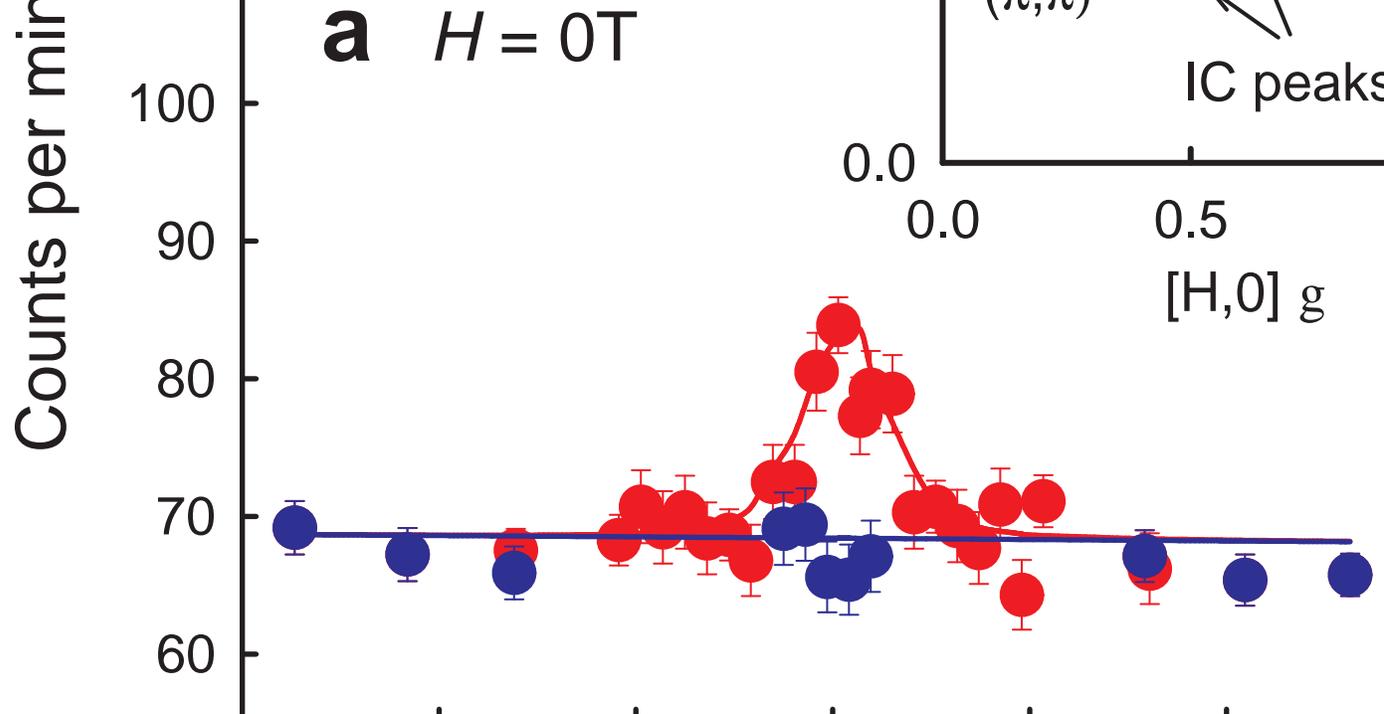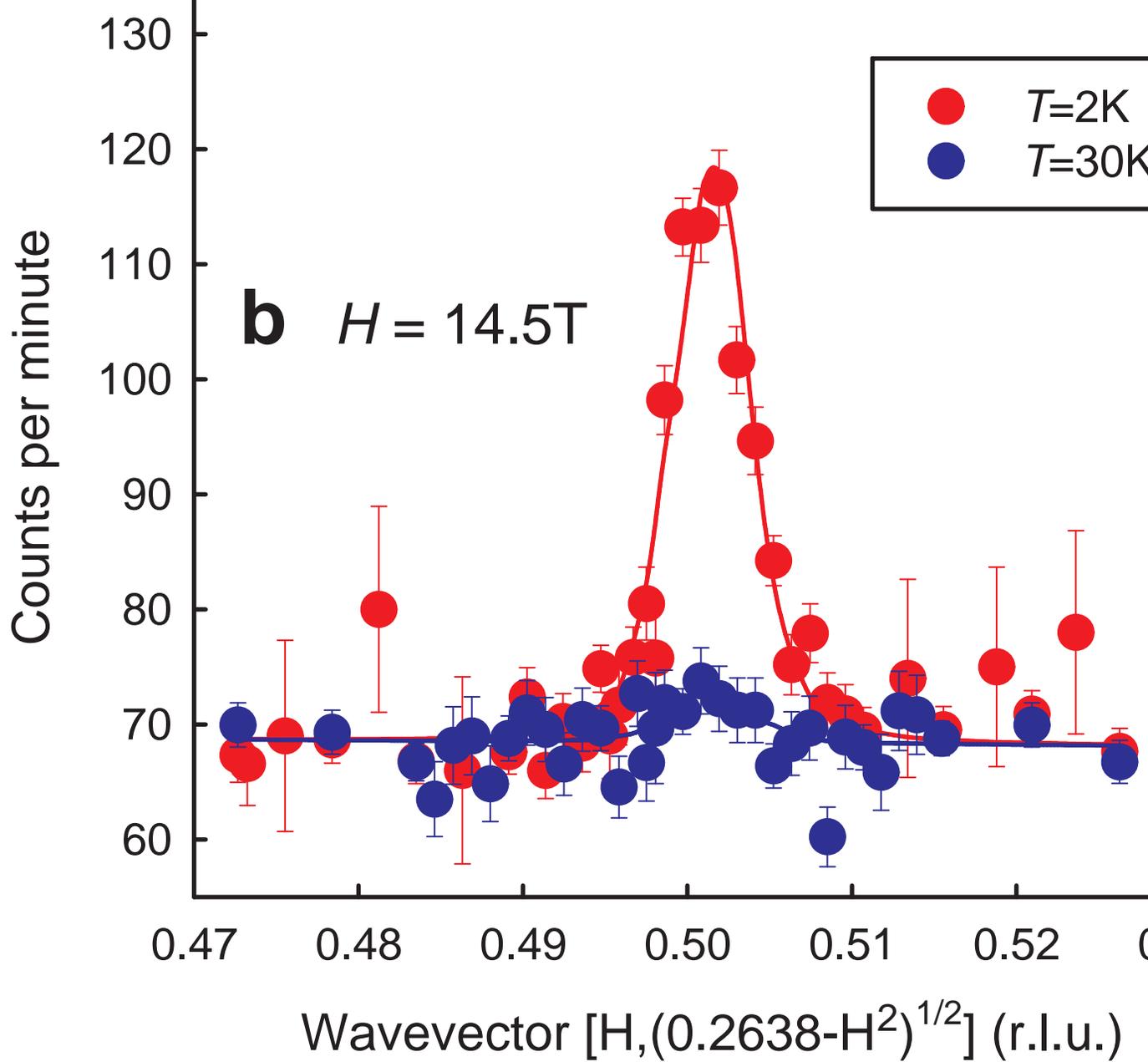

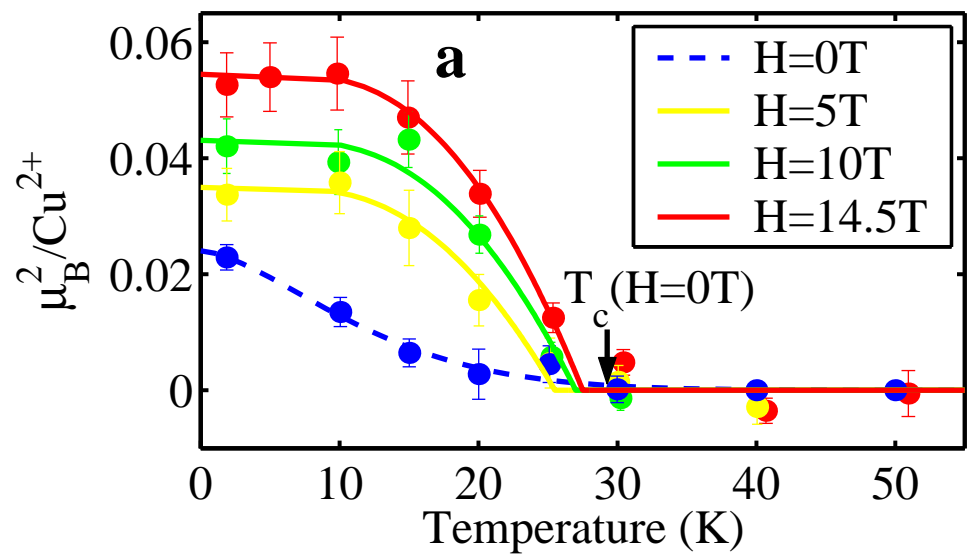
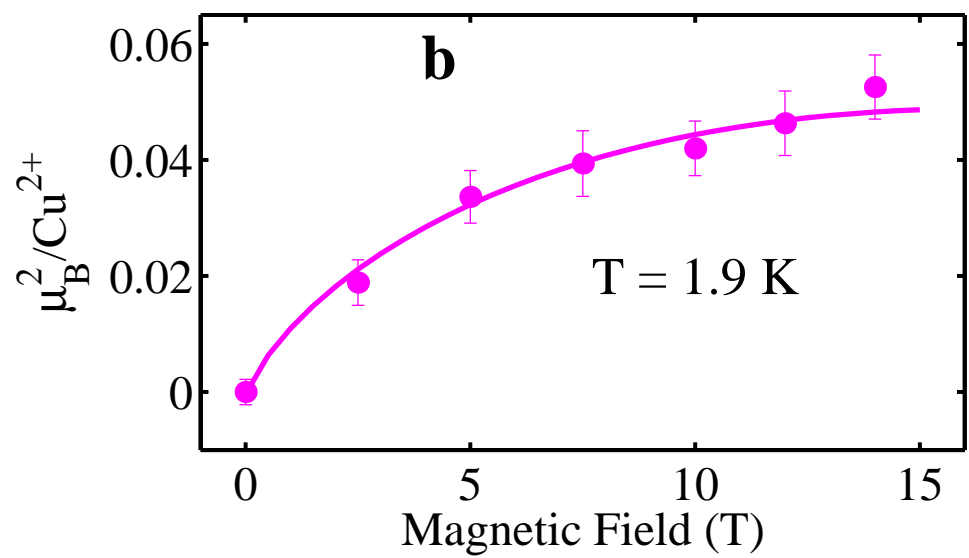